\documentclass[twoside,a4paper]{article}

\usepackage{cite}
\usepackage{amsmath}
\usepackage{amssymb}
\usepackage{graphicx}

\usepackage[T2A]{fontenc}
\usepackage[cp1251]{inputenc}

\usepackage{textcomp}

\usepackage{color}

\usepackage{cmpj2}

\issue{2016}{19}{3}{33002}
\doinumber{10.5488/CMP.19.33002}

\title[Nonlinear model of ice surface softening during friction]%
{Nonlinear model of ice surface softening during friction%
}
\author[A.V.~Khomenko, K.P.~Khomenko, V.V.~Falko]{A.V.~Khomenko\refaddr{label1,label2}\thanks{E-mail: o.khomenko@mss.sumdu.edu.ua}\,, K.P.~Khomenko\refaddr{label1}, V.V.~Falko\refaddr{label1}}
\addresses{
\addr{label1} Sumy State University, 2 Rimskii-Korsakov St., 40007 Sumy, Ukraine
\addr{label2} Peter Gr\"unberg Institut-1, Forschungszentrum-J\"ulich, D-52425 J\"ulich, Germany
}

\authorcopyright{A.V.~Khomenko, K.P.~Khomenko, V.V.~Falko, 2016}
\date{Received September 16, 2015, in final form November 6, 2015}

\begin{document}

\maketitle

\begin{abstract}
The ice surface softening during friction is shown as a result of spontaneous appearance of shear strain caused by external supercritical heating. This transformation is described by the Kelvin-Voigt equation for viscoelastic medium, by the relaxation equations of Landau-Khalatnikov-type and for heat conductivity. The study reveals that the above-named equations formally coincide with the synergetic Lorenz system, where the order parameter is reduced  to shear strain, stress acts as the conjugate field, and temperature plays the role of a control parameter. Using the adiabatic approximation, the stationary values of these quantities are derived. The examination of dependence of the relaxed shear modulus on strain explains the ice surface softening according to the first-order transition mechanism. The critical heating rate is proportional to the relaxed value of the ice shear modulus and inversely proportional to its typical value.
\pacs 05.65.+b, %   Self-organized systems
      05.70.Ln, %   Nonequilibrium and irreversible thermodynamics
      46.55.+d, %   Tribology and mechanical contacts
      62.20.F-, % 	Deformation and plasticity
      64.60.-i, %	General studies of phase transitions
      81.40.Pq %   Friction, lubrication, and wear
\end{abstract}

\section{Introduction}\label{sec:level1}

Ice and snow friction is of great importance in everyday life, sport, nature and industry \cite{review,Persson_book}. The kinetics of ice friction is determined by such processes as adhesion, surface and pressure melting, frictional heating, creep, and fracture \cite{phil_mag_A_2000,JGRB:JGRB17369}. According to \cite{wear_2013}, the question whether temperature or yield stress of ice plays the crucial role at friction is still a subject of some debate. The study \cite{1939} for the first time concludes that the reason for reduced friction
is a water film produced on the ice surface due to frictional heating. Several investigators, in particular \cite{tire:2012,Baurle2007276,jgs_2005}, have extensively developed this idea,  since understanding of liquid film formation conditions is necessary for practical applications. Due to \cite{Farad_Disc_2012}, the premelting layer is formed with fluctuating domains of liquid water and solid ice that resembles the defect structure. The studies of \cite{Kietzig:2010} focused on the thermal conductivity effect that reduces with an increasing sliding velocity.

Consider some aspects of the theoretical models with quantitative estimations. It is generally recognized that ice surface melting during friction does not always take place due to perfect mechanical heating. The results of \cite{tire:2012,Ice-Fric-Akkok-JOT87} explained this feature by thermodynamic arguments. Within the framework of approach \cite{Evans_1976}, the dependencies of friction coefficient for mild steel, Perspex and copper on ice were described regarding the sliding velocity, melting and ambient temperatures, and thermal conductivity of slider. The paper provided the evidence that an enhanced softening of ice above $-2^\circ$C resulted in significant wear. The study \cite{Oksanen_1982} extended the Evans et al. theory \cite{Evans_1976} to the case of hydrodynamic lubrication. Two domains are revealed with different dependencies of friction coefficient on velocity. In the first domain, there is no melting of ice while in the second domain,  friction is governed by the water film at the contact area appearing due to frictional heating. The idea of a prevailing thermal control mechanism during ice friction is confirmed in~\cite{Ice-Fric-Akkok-JOT87}. This survey considers the roles of softening in the wear of rubbing materials, hydrodynamic friction, and squeezing-out of a lubricant film. The investigations \cite{Stiffler_1986,Stiffler_1984} expound the theory for melt lubrication including the case of squeezing-out. This approach, taking into account both hydrodynamic friction and surface roughness, permitted to obtain the expressions for the layer thickness, coefficient of friction and wear. Colbeck \cite{Collbeck_1988} considered that kinetic friction on snow is governed by three components: dry, lubricated, and capillary friction. He constructed several dependencies: 1) the dependence of water film thickness for perfectly insulated and aluminum sliders on temperature; 2)  coefficient of the lubricated friction versus velocity at various temperatures; 3) the dependence of water film thickness on the distance along the lubricated area for plastic and aluminum sliders at various temperatures; 4) total friction versus the length along an aluminum slider at various temperatures. These calculations show that friction force influencing the slider changes slightly over a wide range of velocities and temperatures. This theory is restricted to the suppositions that friction is independent of load, exceptional attention is paid to water friction, approximate estimation of the heat current to the slider, etc.

Recently, there appeared atomistic simulations of ice friction including \cite{Samad_2013,JChemPhys_14,jpc_2009}. Using molecular dynamics method \cite{Samad_2013,JChemPhys_14} and Ginzburg-Landau free energy (Hamiltonian) for the case of first-order phase transition \cite{JChemPhys_14}, it is shown that premelted ice surface film consists of some molecular layers and its thickness grows with load and temperature. This leads to an increased lubrication and to lowered friction due to the weakening of hydrogen bonds between ice molecules. Since the ascent of sliding velocity results in the frictional heating and, eventually, in the growth of temperature and a softened film thickness, the former is maintained  constant due to the introduction of a thermostat in simulations \cite{Samad_2013}. Thus, the calculated friction force vs velocity dependence increases linearly owing to a viscous component of stress arising in the liquid-like film of ice surface during shear.

The relaxation of the shear component of stress proceeds in course of time \cite{upr}
\begin{equation} \tau = \eta /G, \label{1} \end{equation}
where $\eta$ is the dynamical shear viscosity, $G$ is the shear modulus. The principal assumption of \cite{marvan} is as follows: while the kinetic effect of liquid is freezing, the viscosity $\eta$ becomes infinite at a finite shear modulus $G$. However, the situation is opposite to the usual second-order phase transition, where an infinite increase of the $\tau$ at a critical point is also observed. Actually, changing from viscoelastic liquid into a general case, expression (\ref{1}) assumes the form $\tau = \chi / \gamma$, where $\chi$ is the generalized susceptibility, $\gamma$ is the kinetic coefficient [in equation~(\ref{1}) these quantities are $G^{-1}$, $\eta^ {-1}$, respectively] \cite{Landau,kin}. At the phase transition, an infinite increase in susceptibility $\chi$ is observed while the kinetic coefficient $\gamma$ has no singularity.  This is equivalent to the shear modulus $G$ tending to zero at a finite viscosity $\eta$ in equation~(\ref{1}). This situation meets the viscoelastic transition \cite{Bar_Ol,KhYu,pla,Khomenko_TechPhys2007,ftt_2007}.

The underlying assumption of our approach is that ice softening during friction is ensured by self-organization of both stress $\sigma$ and strain $\varepsilon$ shear components, on the one hand, and the temperature $T$, on the other \cite{cmp_2014}. The relationship between components $\sigma$ and $\varepsilon$ is well-known, with the Kelvin-Voigt model describing its simplest case \cite{voigt}. The temperature effect is caused by critical increase in the shear modulus $G (T)$ with a decrease in the temperature: $G=0$ in the water, and $G\not= 0$ in the ice. The governing   equations  are derived in section~\ref{sec:level2} considering the above mentioned circumstances. Section~\ref{sec:level3} presents the study of realization of the conditions of ice surface softening according to the mechanism of continuous second-order transition. The interaction of the mentioned factors results in the onset of the steady state at supercritical value of thermal energy imposed in the surface layer, where the shear strain can take anomalously large values. The lubrication ice friction regime is discussed here, i.e., the model applicable to dry ice friction when the temperature is too low for ice to melt.  Section~\ref{sec:level4} is devoted to the description of ice surface softening by a scheme of discontinuous first-order transition that is observed experimentally in~\cite{JGRB:JGRB17369}. The exposition in this part is basically different from \cite{cmp_2014} since the dependence of a relaxed shear modulus on the strain is of another form (\ref{47a}). Therefore, the resultant figures are quantitatively different from those in \cite{cmp_2014}. Thus, the description of self-organization of adatoms on the semiconductor surface and the ice surface softening within the framework of a similar approach has many differences.

\section{Basic equations}\label{sec:level2}

The configuration for which we present a solution, consists of two rubbing planes of ice or planes of ice and of other material (e.g., solid, rubber and so on), separated by a lubricating softened ice layer. It is widely accepted that the relaxation of the shear component of a strain tensor $\varepsilon$ in the ice surface layer is determined by the Kelvin-Voigt equation for viscoelastic medium \cite{voigt}
\begin{equation}
\dot {\varepsilon}=-\varepsilon/\tau_{\varepsilon}+\sigma/\eta_{\varepsilon}\, ,
\label{7} \end{equation}
where $\tau_{\varepsilon}$ is the Debye relaxation time and $\eta_{\varepsilon}$ is the effective shear viscosity coefficient. The second term on the right-hand side describes the flow of a
viscous liquid due to action of the corresponding shear component of the stress $\sigma$. In stationary state, $\dot\varepsilon = 0$, equation~(\ref{7}) is reduced to the Hooke-type relationship $\sigma = G_{\varepsilon}\varepsilon$, where $G_{\varepsilon} \equiv \eta_{\varepsilon}/\tau_{\varepsilon}\equiv G(\omega)|_{\omega\to 0}$ is the relaxed value of shear modulus ($\omega$ is circular frequency of a periodic external effect).

{Within the framework of the phenomenological Landau theory \cite{Landau,Bar_Ol}, phase transition is governed by free energy $F$ that is expanded  into power series  over $\sigma$ playing the role of an order parameter in study \cite{KhYu}:}
\begin{equation} F = F_0 - \sigma\varepsilon =  \frac{1}{2 G(T)} \sigma^2 + \frac{A}{4}
\sigma^4 - \sigma\varepsilon, \label{l1} \end{equation}
where $G(T)\equiv G(\omega)|_{\omega\to\infty}$ is the non-relaxed shear modulus that depends on the temperature, $\sigma\varepsilon$ implies the external field $\varepsilon$ effect, and $A$ is the positive unharmonicity constant. The equilibrium value of $\sigma$ is determined by the equality
\begin{equation}
\partial F_0/\partial \sigma = \varepsilon, \label{1pre}
\end{equation}
where $F_0$ is free energy at $\varepsilon=0$. The relaxation transition to equilibrium state is described by the Landau-Khalatnikov-type equation \cite{kin,coll1,coll2,KhYu}
\begin{equation} \dot\sigma =
-{\frac{G(T)^2}{\eta}}\left({\frac{\partial F_0}{\partial \sigma}} - \varepsilon \right).
\label{2pre} \end{equation} Here, $\eta$ is a kinetic coefficient, which has the meaning of the shear viscosity. If $\sigma$ is close to its equilibrium value $\sigma_{0}=0$, the linear approximation $\partial F_0 /\partial \sigma \approx \sigma / G(T)$ can be used, where $G(T) \equiv \partial \sigma / \partial\varepsilon = (\partial^2 F_0 / \partial \sigma^2 )^{-1}$. Hence, the relaxation equation (\ref{2pre}) presumes the linear form
\begin{equation}
\tau_\sigma\dot\sigma=-\sigma+G(T) \varepsilon. \label{6} \end{equation}
Here, the first term on the right-hand side describes the relaxation
during time $\tau_{\sigma} \equiv  \eta/G(T)$.
In a steady state $\dot \sigma = 0$, the kinetic equation (\ref{6}) has the form of the Hooke's law
\begin{equation} \sigma = G(T)\varepsilon.  \label{6a}
\end{equation}
Substituting $\varepsilon / \tau_{\sigma}$ for $\partial \varepsilon / \partial t$ in equation~(\ref{6}) reduces it to a Maxwell-type equation for a viscoelastic matter~\cite{upr}.

Note that the effective viscosity $\eta_\varepsilon \equiv \tau_\varepsilon G_\varepsilon$
and a relaxed modulus $G_{\varepsilon} \equiv \eta_\varepsilon/\tau_{\varepsilon}$ do not
coincide with the real viscosity $\eta$ and non-relaxed modulus $G(T)$, respectively. This is caused by a different physical meaning of the Landau-Khalatnikov-type (\ref{6}) and
the Kelvin-Voigt (\ref{7}) equations \cite{voigt,upr,KhYu}. The values $G_\varepsilon\,$, $\eta$, $\eta_\varepsilon$ very weakly depend  on the ice surface layer temperature $T$, while the shear modulus $G(T)$ vanishes when the temperature decreases to $T_{\text c}$ \cite{marvan,book_Intech,glass1,glass2}. Further, the temperature dependencies are used for the  approximation:
$G_{\varepsilon}(T)$, $\eta(T)$, $\eta_\varepsilon(T) = \text{const}$,
\begin{equation}
G(T) = G_0\left(T/T_{\text c}-1\right), \label{8}
\end{equation}
where $G_{0} \equiv G(T=2T_{\text c})$ is a typical value of modulus.

In order to present the self-organization process \cite{Haken,zhetph,KhYu,glass1,glass2,physa_soc,jam_pre}, the kinetic equation for temperature $T$ is needed for completing the equations system (\ref{7}) and (\ref{6}), which contains the order parameter $\varepsilon$, conjugate field $\sigma$, and control parameter $T$. Employing the  approach \cite{cmp_2014}, based on the elasticity theory relationships in~\cite{upr}, $\S~32$, the following equation can be derived:
\begin{equation}
c_p\dot T = \kappa\nabla^2T - \sigma\varepsilon/\tau_{\varepsilon}+\sigma^2/\eta_{\varepsilon}\, , \label{17gl_a}
\end{equation}
where $c_p$ is the heat capacity, $\kappa$ is the heat conductivity. The last term on the right-hand side stands for dissipative heating of a viscous liquid, flowing under the effect of the stress $\sigma$, that can be neglected in the case under consideration. {On the other hand, the one-mode approximation $(\kappa /l^2) (\tau _{T}Q-T)\approx \kappa\nabla^2T$ can be used with acceptable accuracy in equation~(\ref{17gl_a}) \cite{upr,KhYu,glass1,jtph_10_8,Khomenko_UJP2009,cmp_2006}.} Thus, we consider the thermal effect of friction surfaces whose value is not reduced to the Onsager component and is fixed by external conditions ($l$ is the scale of heat conductivity, i.e., the distance into which heat penetrates ice, $\tau_{T}\equiv l^2 c_p / \kappa$ is the time of heat conductivity):
\begin{eqnarray}
\label{17gl}
\tau _{T}\dot T &=& (\tau _{T}Q-T) - \frac{l^2\sigma\varepsilon} {\kappa \tau_{\varepsilon}}\,,\\
%\end{equation}
%\begin{eqnarray}
Q &=& Q_0 + \sigma^2/c_p \eta_\varepsilon\,.
\label{14d}
\end{eqnarray}
Here, $Q_0$ is a heat flow from the surrounding solids to the surface layer. The square contribution of the stress is implied to be included in the rubbing surfaces temperature $T_{\text{e}} = \tau _{T}Q$. The obvious account of this term leads to a significant complication of the subsequent analysis, though it results only in renormalization of the quantities. Therefore, for our further consideration, the  component $T_{\text{e}}= \tau _{T}Q$ in equation~(\ref{17gl}) is presumed to be constant. {It is noteworthy that during derivation of equation~(\ref{17gl}) we accepted the equilibrium value of the temperature of ice surface layer $T_{00}$ to be equal to zero.  Evidently, contrary to the ice surface being heated initially to the temperature $T_{00} \not=0$, the term $T_{00}/\tau_T$ should enter equation~(\ref{14d}). This term describes the relaxation of the current temperature of ice surface layer to its equilibrium value~$T_{00}$ in the absence of the heat flow $Q$ from the background solids.}

It is convenient to introduce the following measure units:
\begin{equation}
\sigma_{\text{s}}=\left(c_p\eta_\varepsilon T_{\text c}/\tau _{T}\right)^{1/2},
\qquad \varepsilon_{\text{s}}= \sigma_{\text{s}}/G_\varepsilon\, ,\qquad T_{\text c}
\label{10} \end{equation}
for the variables $\sigma$, $\varepsilon $, $T$, respectively. Then, the basic equations (\ref{7}), (\ref{6}), and (\ref{17gl}) are reduced to a form applicable to any viscoelastic medium \cite{cmp_2014}:
\begin{eqnarray}
\tau_{\varepsilon }\dot{\varepsilon}&=&-\varepsilon + \sigma ,
\label{11} \\
\tau_{\sigma}\dot{\sigma}&=&-\sigma +g(T-1)\varepsilon , \label{12} \\
\tau _{T}\dot{T}&=&(\tau _{T}Q-T) - \sigma \varepsilon,
\label{13} \end{eqnarray} where the constant
\begin{eqnarray} g=\frac{G_0}{G_\varepsilon} \label{14} \end{eqnarray} is introduced. The equations (\ref{11}) -- (\ref{13}) have a form similar to the Lorenz scheme \cite{Haken} which allows us to denote the thermodynamic phase and kinetic transitions \cite{zhetph,KhYu,glass1,glass2,physa_soc,jam_pre}.

\section{Continuous second-order transition}\label{sec:level3}

In general, equations~(\ref{11}) -- (\ref{13}) have no analytical solution, therefore, the adiabatic approximation is used for this purpose:
\begin{equation}
\tau_{\sigma} \ll \tau_{\varepsilon},\qquad \tau_{T} \ll \tau _{\varepsilon}\, .
\label{15} \end{equation}
This approach suggests that in the course of evolution, stress $\sigma(t)$ and temperature $T(t)$ follow the variation of strain $\varepsilon(t)$. {The minimal relaxation time of strain $\tau_\varepsilon$ is defined by time of reorientations of the water molecules at the freezing point of fresh water $2\times 10^{-5}$~s and $\tau_\varepsilon$ increases by several orders of magnitude at confinement of premelted ice layer between the rubbing surfaces \cite{water_book,Lang_2008}.} The microscopic Debye time is estimated by relation $\tau_{\sigma}\approx a/c {\sim} 10^{-12}$~s, where $a\sim 1$~nm is the lattice constant or intermolecular distance, and $c\sim 10^3$~m$/$s is the sound velocity. Therefore, the first of inequalities (\ref{15}) is valid. The second condition (\ref{15}) can be written in the form
\begin{equation} l \ll L,
\label{e} \end{equation}
where the maximal value of the characteristic length of heat conductivity
\begin {equation}
L = \sqrt{\chi\nu_\varepsilon \over c_\varepsilon^2}\, , \label{f} \end{equation}
the thermometric conductivity $\chi \equiv \kappa /c_p\,$, the effective kinematic viscosity $\nu_\varepsilon \equiv \eta_\varepsilon / \rho$, and the sound velocity $c_\varepsilon \equiv(G_\varepsilon/\rho)^{1/2}$
are introduced ($\rho$ is the medium density). For ice $\rho \approx 916$~kg$/$m$^3$, $\kappa \approx 2.22$~W$/$m$\cdot$K, $c_p \approx 2050$~J$/$kg$\cdot$K, $G_\varepsilon \approx 10~\text{GPa}$ and the water dynamical shear viscosity $\eta_\varepsilon \approx 1.8\cdot 10^{-3}$~Pa$\cdot$s at the $T = 0^\circ\text{C}$, the value of $L\approx 10$~nm that agrees with the  experiments \cite{Baurle2007276,jgs_2005} as well as with the field and atomistic theories \cite{JChemPhys_14,jpc_2009}.

Then, equaling the left-hand sides of equations~(\ref{12}) and (\ref{13}) to zero, we can express stress $\sigma$ and temperature $T$ in terms of strain $\varepsilon$:
\begin{eqnarray}
\sigma &=& \frac{g\varepsilon \left(T_{\text{e}} -1 \right)}{1+g\varepsilon^{2} }\, , \label{16} \\
T&=&1+\frac{T_{\text{e}} -1}{1+g\varepsilon^{2}}\, .  \label{17}
\end{eqnarray}
According to equation~(\ref{17}), at the important interval of the parameter $T_{\text{e}}=\tau _{T}Q>1$ values, the temperature $T$ decreases monotonously with strain growth $\varepsilon$ from the value
$T_{\text{e}}$ at $\varepsilon = 0$ to $(T_{\text{e}}+1)/2$ at $\varepsilon =\varepsilon_{\text{m}}\equiv\sqrt{1/g}$. Obviously, the negative feedback of the stress and strain on the temperature in equation~(\ref{13}) leads to this descent. Such an activity is connected with implementation of the Le Chatelier principle for this problem. Indeed, the positive feedback of strain and temperature on stress in equation~(\ref{12}) is the reason for ice melting. Consequently,  the self-organization should occur  more intensively with the increase in temperature. However, in accordance with equation~(\ref{13}), the system demonstrates such a pattern that the consequence of transition, i.e.,
ascent of strain, is a drop of temperature, whose growth is the reason for $\varepsilon$ increasing. The stress vs strain dependence~(\ref{16}) has the linear Hooke's section at $\varepsilon \ll \varepsilon_{\text{m}}$ with the effective shear modulus $G_{\text{ef}} \equiv g\left(T_{\text{e}} -1 \right)$. At $\varepsilon = \varepsilon_{\text{m}}\,$, the function $\sigma (\varepsilon)$ rises to a maximum and at $\varepsilon > \varepsilon_{\text{m}}$ it reduces, which has no physical meaning. Thus, the constant $\varepsilon_{\text{m}}\equiv\sqrt{1/g}$ corresponds to maximal strain. The growth of a typical value of the modulus $G_0$ decreases the maximal strain $\varepsilon_{\text{m}}$ and increases the effective modulus $G_{\text{ef}}$ whose value is proportional to the background ice temperature $T_{\text{e}}\,$.

Substitution of equation~(\ref{16}) into equation~(\ref{11}) produces the
Landau-Khalatnikov-type equation~\cite{kin,coll1,coll2,PhysRevLett_Metlov}
\begin{equation}
\tau_{\varepsilon}\dot{\varepsilon}=-\partial V/\partial \varepsilon. \label{18}
\end{equation}
Here, the synergetic potential has the form
\begin{equation}
V=\frac{1}{2}\left[\varepsilon^{2}+\left(1-T_{\text{e}}\right)\ln \left(1+g\varepsilon^{2}\right)\right],  \label{19}
\end{equation}
that is reduced after expansion of logarithm over $\varepsilon^2$ to free energy used in~\cite{Bar_Ol} for a description of viscoelastic transition of unstructured condensed matter.
At stationary state $\dot{\varepsilon}=0$, the potential (\ref{19}) acquires a minimum. When the temperature $T_{\text{e}}$ is lower than the critical value
\begin{equation}
T_{\text{c}0}=1+g^{-1},\qquad g\equiv G_{0}/G_\varepsilon < 1,\qquad G_\varepsilon \equiv \eta_\varepsilon/\tau _{\varepsilon}\, , \label{20}
\end{equation}
this minimum corresponds to $\varepsilon=0$, i.e., the ice surface is not softened. In the opposite situation $T_{\text{e}}>T_{\text{c}0}\,$, the steady shear strain acquires a nonzero value
\begin{equation}
\varepsilon_{0}=\left[T_{\text{e}}-(1+g^{-1})\right]^{1/2} \label{21}
\end{equation}
increasing with $T_{\text{e}}$ growth in accordance with the root law. This causes the ice softening. Consideration of strain proportionally to the thickness of premelting ice layer reveals a qualitative agreement of equation~(\ref{21}) with the results of molecular dynamics simulations and statistical field theory \cite{Samad_2013,JChemPhys_14}. Besides, in line with \cite{Samad_2013}, we reckon that the friction reduces with the temperature growth because hydrogen couplings fail.

Using equations~(\ref{16}) and (\ref{17}) we get the stationary values of stress and temperature:
\begin{equation}
\sigma_{0}=\varepsilon_{0}\, , \qquad T_{0}=1+g^{-1}. \label{22}
\end{equation}
It is noteworthy that, on the one hand, the steady temperature $T_{0}$
coincides with the critical value (\ref{20}) and, on the other hand,
its value differs from the temperature $T_{\text{e}}\,$. Since $T_{\text{c}0}$ is the minimal temperature at which the ice softening proceeds, this statement represents the effect of a negative feedback of stress $\sigma$ and strain $\varepsilon$ on temperature $T$ [see the last term on the right-hand side of equation~(\ref{13})]. Self-organization occurs in the limit because the sample temperature drops so much. At a stationary state, the non-relaxed shear modulus is equal to the relaxed one
\begin{equation} G_{\text{s}} = G_{\varepsilon}\, . \label{23} \end{equation}
Thus, the surface softening is represented in the model by increasing $\varepsilon$ and $T_{\text{e}}$ because the ice modulus is a fixed quantity.

Two cases can be distinguished by the parameter $g=G_{0}/G_\varepsilon\,$.
At $g\gg 1$, that meets the large value of the modulus $G_{0}\,$, equations~(\ref{20})--(\ref{22}) assume the forms
\begin{equation} \varepsilon_{0}=(T_{\text{e}}-1)^{1/2}, \qquad T_{0}=T_{\text{c}0}=1, \label{24}
\end{equation} corresponding to the ``ice (brittle)'' limit. The reverse case $g\ll 1$ (small modulus $G_{0}$) corresponds to the ``strongly viscous liquid''
\begin{equation} \varepsilon_{0}=(T_{\text{e}}-g^{-1})^{1/2},\qquad T_{0}=T_{\text{c}0}=g^{-1}=G_\varepsilon/G_{0}\, .
\label{25} \end{equation}

\section{Discontinuous first-order transition due to deformational defect of modulus}\label{sec:level4}

By using the Kelvin-Voigt equation (\ref{7}), we suggest the validity of the idealized
Genki model for the stress $\sigma$ vs strain $\varepsilon$ dependence.
It implies the realization of the Hooke's law $\sigma = G_\varepsilon\varepsilon$ at $\varepsilon < \varepsilon_{\text{m}}$ and the constant $\sigma_{\text{m}}=G_\varepsilon\varepsilon_{\text{m}}$ at $\varepsilon\geqslant\varepsilon_{\text{m}}$ [$\sigma_{\text{m}}$, $\varepsilon_{\text{m}}$ are the maximal stress and strain, $\sigma > \sigma_{\text{m}}$ results in a viscous flow with the
deformation rate $\dot\varepsilon = \left (\sigma - \sigma_{\text{m}}\right)
/\eta_\varepsilon$]. Actually, the $\sigma(\varepsilon)$ dependence possesses two regions: the first one, Hookean, has a large tilt corresponding to the relaxed shear modulus $G_\varepsilon\,$, then follows a gentler sloping section of the plastic deformation whose slope is fixed by the hardening factor $\Theta < G_\varepsilon\,$. Indeed, the described case means the relaxed shear modulus in equation~(\ref{7}) depending on the strain value. For example, the simplest approximation is used in \cite{Khomenko2007165,jfd_2007} for the case of ultrathin lubricant film melting
\begin{equation} G_\varepsilon (\varepsilon) = \Theta + {G_\varepsilon - \Theta \over 1 +
\left( \varepsilon /\varepsilon_p \right)^\beta }\, , \label{47}
\end{equation}
which describes the above mentioned transition from the elastic deformation mode to the plastic one ($\beta$~is the positive constant). {It takes place at a characteristic value of the strain $\varepsilon_p\,$.} It is noteworthy that a relationship of equation~(\ref{47}) type has been proposed, for the first time, by Haken \cite{Haken} in order to represent the rigid mode of laser radiation. This equation was used for denoting the first-order phase transition \cite{zhetph,physa_soc,jam_pre,KhYu,glass1,glass2}. Experimental dependencies of shear force on displacement for friction of ice on ice demonstrate the similar peculiarities \cite{JGRB:JGRB17369} but, as a rule, the plastic section is horizontal. To describe such a behavior, it is necessary to surmise that $\Theta = 0$ and $\beta = 1$ in (\ref{47}):
\begin{equation} G_\varepsilon (\varepsilon) = \frac {G_\varepsilon} {1 + \varepsilon /\varepsilon_p}\, . \label{47a}
\end{equation}
Moreover, the interpretation of ice surface premelting as a plastic act due to first-order transition agrees with the results of atomistic and statistical field approaches, developed in studies~\cite{Samad_2013,JChemPhys_14}.
\begin{figure}[!t]
\centerline{
\includegraphics[width=0.25\textwidth]{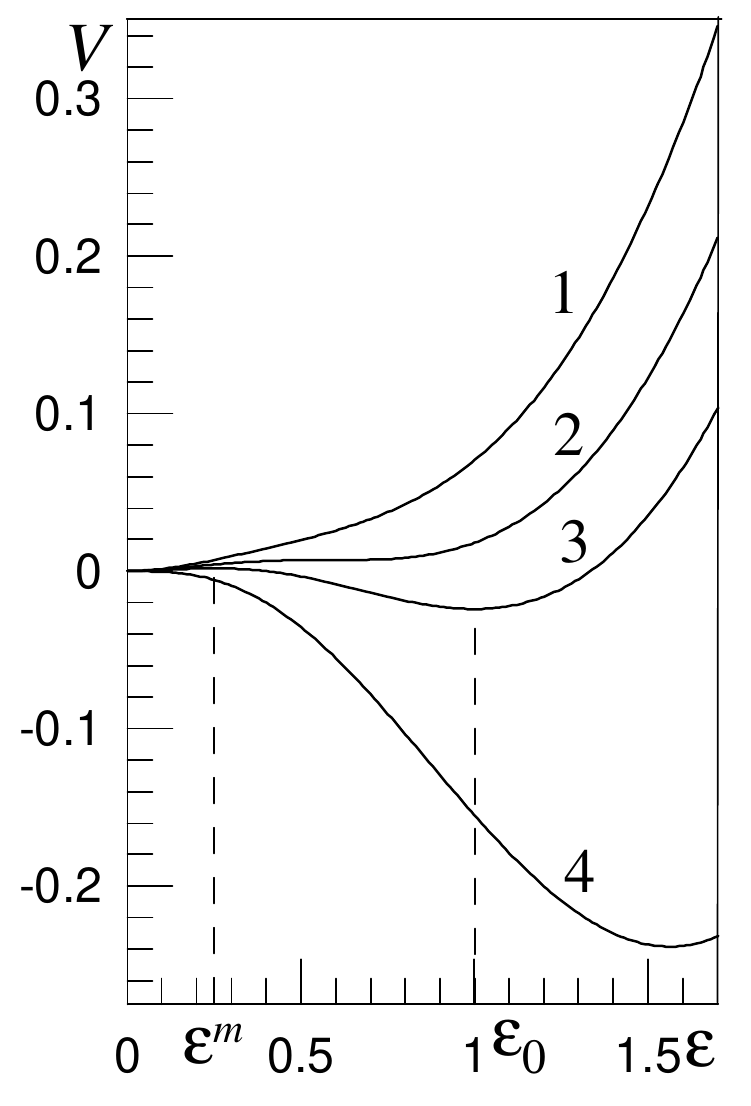}
}
\caption{\label{fig1} Dependence of the synergetic potential on the strain at $g {=} 0.8$, $\alpha {=} 0.8$ and different temperatures: (curve~1) $T_{\text{e}} {<} T_{\text{c}}^0$, (curve 2) $T_{\text{e}}{=}T_{\text{c}}^0$, (curve 3) $T_{\text{c}}^0 {<} T_{\text{e}} {<} T_{\text{c0}}\,$, and (curve 4) $T_{\text{e}}{\geqslant} T_{\text{c}0}\,$.}
\end{figure}
\begin{figure}[!b]
\centerline{
\includegraphics[width=0.4\textwidth]{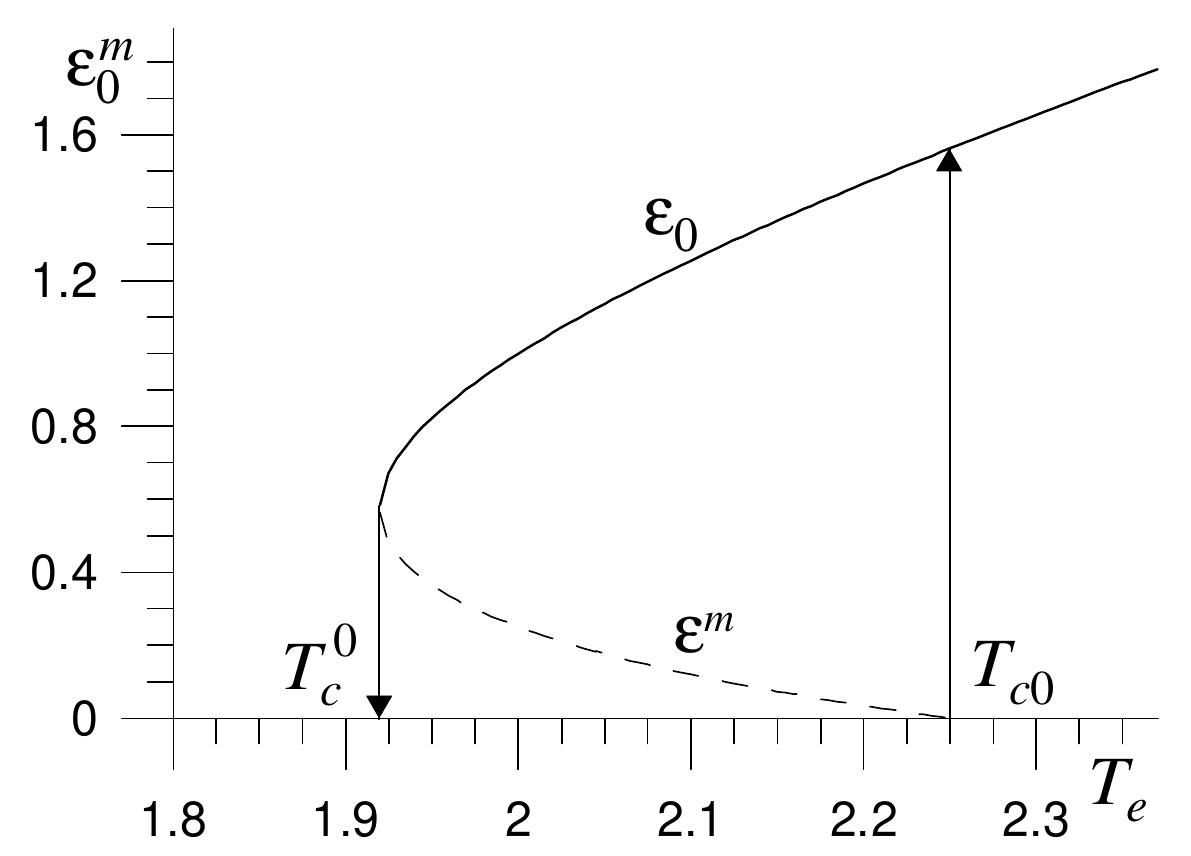}
}
\caption{\label{fig2} Dependence of the steady strains on the temperature $T_{\text{e}}$ at parameters of figure~\ref{fig1} (the solid curve meets the stable value $\varepsilon_0\,$, the dashed curve  corresponds to the unstable one, $\varepsilon^{\text{m}}$).}
\end{figure}

Using the adiabatic approximation (\ref{15}) for the Lorenz equations (\ref{11}) -- (\ref{13}), where $G_\varepsilon$ is replaced by the dependence $G_\varepsilon (\varepsilon)$ (\ref{47a}), the Landau-Khalatnikov equation (\ref{18}) is derived. The synergetic potential
\begin{eqnarray}
V=\frac{1}{2}\varepsilon^{2} {+} (1{-}T_{\text{e}})\left\{\frac{1}{2}\ln\left( 1{+}g\varepsilon^2 \right) {+} \frac{1}{\alpha} \left[\varepsilon {-} \frac{1}{\sqrt{g}} \arctan\left( \sqrt{g}\varepsilon \right)\right]\right\}
\label{51} \end{eqnarray}
differs from (\ref{19}) by the last term containing the constant $\alpha {\equiv} \varepsilon_p / \varepsilon_{\text{s}}\,$. At a minor value of temperature $T_{\text{e}}\,$, the dependence~(\ref{51}) is monotonously increasing with a minimum at $\varepsilon = 0$ corresponding to stationary state of ice (curve~1 in figure~\ref{fig1}). As shown in figure~\ref{fig1}, plateau (curve~2) appearing at
\begin{equation} T_{\text{c}}^0 {=} 1 {-} 2\alpha^{2} {+} 2\alpha \sqrt{\alpha^{2} {+} g^{-1}}, \label{52}
\end{equation}
for $T_{\text{e}} > T_{\text{c}}^0$ is transformed into a minimum at the strain $\varepsilon_0\not= 0$, and maximum at $\varepsilon^{\text{m}}$ that separates the minima at the values $\varepsilon = 0$ and $\varepsilon = \varepsilon_0$ (curve~3). With the further growth of the temperature $T_{\text{e}}\,$, the ``ordered'' phase minimum, corresponding to the ice softened structure $\varepsilon = \varepsilon_0\,$, becomes deeper, and the height of the interphase barrier diminishes, disappearing at the critical value $T_{\text{c}0}=1+g^{-1}$ (\ref{20}). The stationary values of strain have the form (see figures~\ref{fig1}~and~\ref{fig2})
\begin{eqnarray}
\varepsilon_0^{\text{m}} = \left(2\alpha\right)^{-1} \left\{
\left(T_{\text{e}}-1\right) {\pm} \sqrt{\left(T_{\text{e}}{-}1\right)^{2}{-}4g^{-1}\alpha^{2} \left[1{-}g\left(T_{\text{e}}{-}1 \right) \right] } \right\}, \label{55} \end{eqnarray}
where the upper sign indicates the stable softened ice structure and the lower sign denotes the unstable one. At $T_{\text{e}}\geqslant T_{\text{c}0}\,$, the shape of the $V$ vs $\varepsilon$ dependence is similar to the one at the absence of the deformational modulus defect (see curve~4 in figure~\ref{fig1}). {With a decrease of the rubbing surfaces temperature~$T_{\text{e}}$ the interphase barrier disappears at its critical value $T_{\text{c}}^0$. At the same time, the potential minimum, corresponding to the ice softened structure $\varepsilon = \varepsilon_0\,$, vanishes too and this phase transforms into solid ice meeting the minimum at $\varepsilon = 0$. It is noteworthy that temperature $T_{\text{c}}^0$ represents the low limit of the range of realization of first-order phase transition $T_{\text{c}}^0<T_{\text{e}}<T_{\text{c}0}\,$. At the point $T_{\text{c}}^0\,$, stable $\varepsilon = \varepsilon_0$ and unstable $\varepsilon = \varepsilon^{\text{m}}$ solutions of stationary states equation $\partial V(\varepsilon) / \partial\varepsilon = 0$ are equal $\varepsilon_0 = \varepsilon^{\text{m}}$.}

\begin{figure}[!t]
\centerline{
\includegraphics[width=0.35\textwidth]{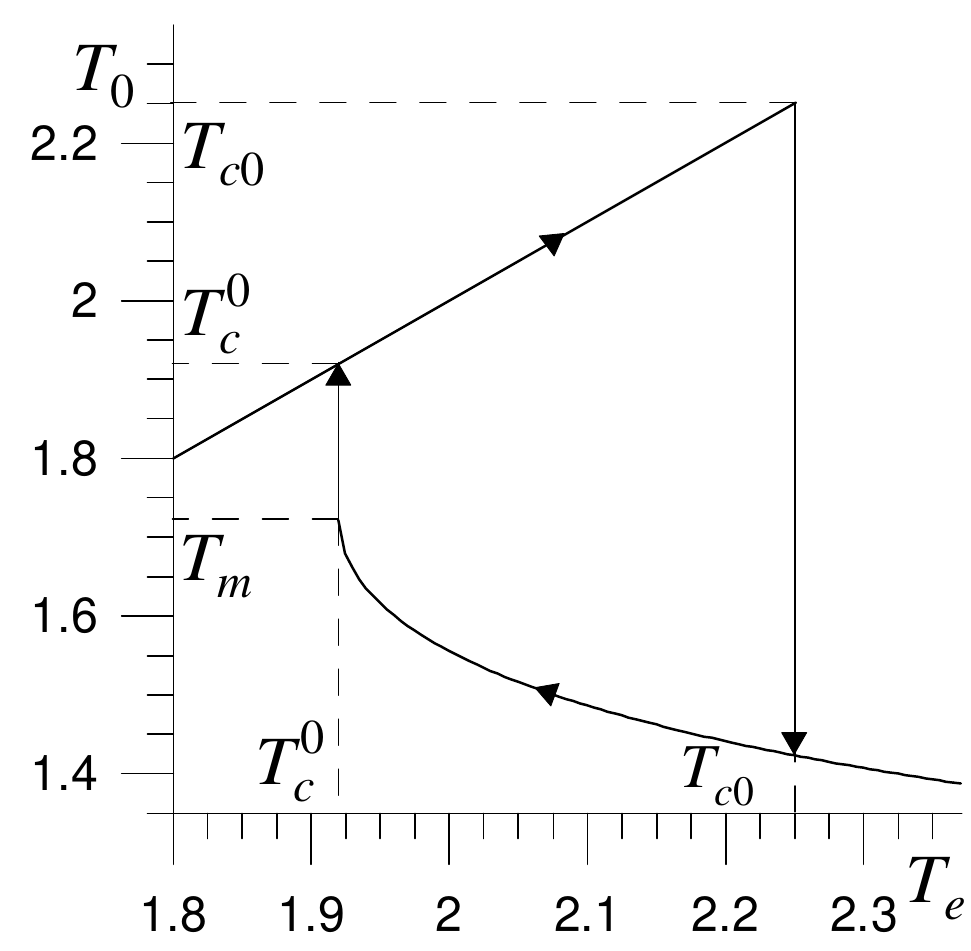}
}
\caption{\label{fig3} Dependence of the stationary temperature $T_0$ of the softened ice layer on the temperature $T_{\text{e}}$ at parameters of figure~\ref{fig1}.}
\end{figure}

The potential barrier, typical of a synergetic first-order transition, manifests itself  only at the deformational defect of the modulus. Since it always takes place \cite{JGRB:JGRB17369,JChemPhys_14,Khomenko2007165,jfd_2007}, the studied ice softening represents the  synergetic first-order transition. This occurrence is more complex than the thermodynamic phase transition. Obviously, in the latter case, the stationary value of the softened layer temperature
$T_0$ coincides with a thermostat value $T_{\text{e}}\,$. In this investigation, the $T_0$ is equal to the critical value $T_{\text{c}0}$ for the synergetic second-order transition (see section~\ref{sec:level3}). When the modulus defect is examined, the temperature
\begin {equation}
T_0=1+\frac{T_{\text{e}} -1}{1+g\varepsilon_0^{2}}\, ,
\label{56}
\end{equation}
corresponding to the minimum of the dependence (\ref{51}), presents itself. In accordance with equations~(\ref{55}) and (\ref{56}), the magnitude $T_0$ smoothly descends from the value
\begin{eqnarray}
T_{\text{m}}=1+\frac{T_{\text{c}}^0 - 1}{1+g\left(\varepsilon_0^{\text{c}}\right)^{2}}, \qquad \varepsilon_0^{\text{c}} =  \frac{T_{\text{c}}^0-1}{2\alpha}\,, \label{58} \end{eqnarray}
at $T_{\text{e}} = T_{\text{c}}^0$, to $1$ at $T_{\text{e}} \to \infty$. As delineated in figure~\ref{fig3}, the steady temperature $T_0$ grows linearly from $0$ to $T_{\text{c}0}\,$, with $T_{\text{e}}$ being in the same interval and, after the drop at $T_{\text{e}}=T_{\text{c}0}\,$, the value $T_{0}$ smoothly decreases. If the temperature $T_{\text{e}}$ then descents, the steady-state temperature $T_0$ increases. At the point $T_{\text{c}}^0$ [equation~(\ref{52})], temperature $T_0$ jumps from the $T_{\text{m}}$ [equation~(\ref{58})] up to the  $T_{\text{c}}^0$.
For $T_{\text{e}} < T_{\text{c}}^0$, again the steady temperature $T_0$ coincides with $T_{\text{e}}\,$.

The maximal softened ice temperature (\ref{58}) is lower than the minimal temperature of the background ice $T_{\text{c}}^0 > 1$ (\ref{52}). As depicted in figure~\ref{fig3}, at $T_{\text{e}} > T_{\text{c}}^0$, the steady-state temperature $T_0$ of the ice surface layer is less than the $T_{\text{e}}\,$.

\section{Summary}\label{sec:level5}

This consideration shows that the ice surface softening during friction is conditioned by the self-organization of the strain and stress shear components, on the one hand, and by the layer temperature, on the other hand. At this, strain $\varepsilon$ is the order parameter, stress $\sigma$ plays the role of the conjugate field, and temperature $T$ acts as the control parameter. The positive feedback of $T$ and $\varepsilon$  on $\sigma$ [see equation~(\ref{12})] leads to self-organization. The temperature dependence of shear modulus in equations~(\ref{6}) and (\ref{8}) has a crucial role. The assumption about shear modulus vs strain dependence allows us to acquire the relationships for temperatures of absolute instability of the softened ice layer $T_{\text{c}}^0$ [equation~(\ref{52})] and  stability limit of the solid ice $T_{\text{c}0}$ [equation~(\ref{20})]. The real temperature of transition, lying in the $(T_{\text{c}}^0, T_{\text{c}0})$ range, can be extracted from the equality $V(0)=V(\varepsilon_0)$ of potentials in various phases. The analysis of equation~(\ref{20}) demonstrates that the softening begins earlier in the systems with a large typical $G_0$ and minor relaxed $G_\varepsilon$ values of shear modulus. The kinetic Landau-Khalatnikov equation (\ref{18}) with the synergetic potential (\ref{51}) describes this first-order transition. The freezing  of softened ice at $\tau_\varepsilon = \infty$ can occur ($\dot\varepsilon\to 0$) even in a nonstationary state $\partial V /\partial\varepsilon \not= 0$.

The obtained expressions for temperatures (\ref{20}) and (\ref{52}) are of an engineering character because they can be used to predict the friction reduction or increase as well as to remove the negative effect of the interrupted mode of ice friction {being the main reason for destruction of the rubbing surfaces}. At this temperature interval, the stick-slip mode of friction is possible, characterized by transitions between two dynamic states during the stationary sliding. The latter is observed due to the presence of rapidly fluctuating (in space and time) domains of ice and liquid-like ice \cite{Farad_Disc_2012,pla,Khomenko_TechPhys2007,Khomenko_UJP2009}. In the coming work, we are aiming to theoretically define the parameters of the rubbing surfaces at which the ice surface layer is in a liquid-like state and the friction between the surfaces decreases or increases. It is planned to combine the contact mechanics, thermodynamic and nonlinear models in order to define the behavior of the system at intermediate velocity interval, where the frictional heating leads to the lowering of friction by either thermal softening of a thin surface layer, or by the formation of nonuniform thin surface film~\cite{Persson_book}. By joint use of these approaches, it is possible to construe why the ice friction sharply diminishes with ascent of shear rate before reaching the velocity at which a thin homogeneous water layer appears on the ice surface at the expense of frictional heating.

\section*{Acknowledgements}

The work was supported by the grants of the Ministry of Education and Science of Ukraine ``Nonequilibrium thermodynamics of metals fragmentation and friction of spatially nonhomogeneous boundary lubricants between surfaces with nanodimensional irregularities'' (No.~0115U000692) and for a research visit to the Forschungszentrum J\"{u}lich (Germany). A.V.K. is grateful to Dr.~Bo~N.J.~Persson for kind  invitation, hospitality, initiation of this work and fruitful discussions during his stay at the Forschungszentrum J\"{u}lich (Germany). A.V.K. expresses gratitude to him and to the organizers of the conference ``International Conference on Friction and Energy Dissipation in Man-Made and Biological Systems'' (Novem\-ber~5--8, 2013, Miramare, Trieste, Italy) for invitation and sponsorship of participation.

%\bibliographystyle{cmpj}
%\bibliography{ice_AFM_cmp}

\ukrainianpart

\title{Нелінійна модель розм'якшення поверхні льоду при терті}
\author{О.В.~Хоменко\refaddr{label1,label2}, К.П.~Хоменко\refaddr{label1}, В.В.~Фалько\refaddr{label1}}
\addresses{
\addr{label1} Сумський державний університет, вул. Римського-Корсакова, 2, 40007 Суми, Україна
\addr{label2} Інститут Петера Грюнберга-1, Дослідницький центр Юліха, 52425 Юліх, Німеччина
}

\makeukrtitle

\begin{abstract}
\tolerance=3000%
Розм'якшення поверхні льоду при терті показано як результат спонтанної появи деформації зсуву, викликаної зовнішнім надкритичним нагрівом. Це перетворення описується рівнянням Кельвіна-Фойгта для в'язкопружного середовища, релаксаційними рівняннями типу Ландау-Халатнікова і теплопровідності. Показано, що вказані рівняння формально збігаються з синергетичною системою Лоренца, де параметр порядку зводиться до деформації зсуву, напруження є спряженим полем, і температура відіграє роль керувального параметра. При використанні адіабатичного наближення отримані стаціонарні значення цих величин. Розгляд залежності релаксованого модуля зсуву від деформації пояснює розм'якшення поверхні льоду згідно з механізмом переходу першого роду. Критична швидкість нагріву пропорційна значенню релаксованого модуля зсуву льоду і обернено пропорційна його характерному значенню.
\keywords фазовий перехід, реологія, пластичність, деформація, напруження, модуль зсуву

\end{abstract}

\end{document}